\documentclass[12pt]{article}

\usepackage[colorlinks,citecolor=blue,urlcolor=black]{hyperref}
\usepackage{bm}
\usepackage{amssymb}
\usepackage{amsmath}
\usepackage{amsthm}
\usepackage{mathrsfs}
\usepackage{dsfont}
\usepackage{thmtools} 
\usepackage{thm-restate}
\usepackage{booktabs}
\usepackage{multirow}
\usepackage{diagbox}
\usepackage{adjustbox}
\usepackage{xcolor}
\usepackage{natbib}
\usepackage{url}
\usepackage{appendix}
\usepackage[doublespacing]{setspace}
\usepackage[margin=1in]{geometry}
\usepackage{empheq}  %equations with left brace

\usepackage{enumerate}

\def\mathR{\mathbb{R}}

\newcommand{\blind}{1}

\begin{document}

\def\spacingset#1{\renewcommand{\baselinestretch}%
{#1}\small\normalsize} %\spacingset{1}

\if1\blind
{
	\title{Joint Modeling and Prediction of Massive Spatio-Temporal Wildfire Count and Burnt Area Data with the INLA-SPDE Approach}
	\author{Zhongwei Zhang, Elias Krainski, Peng Zhong, H\r{a}vard Rue,\\ and Rapha\"el Huser \\
	Statistics Program, CEMSE Division, King Abdullah University of \\ Science and Technology}
	\date{}
	\maketitle
} \fi

\if0\blind
{
	\bigskip
	\bigskip
	\bigskip
	\begin{center}
		{\LARGE\bf Title}
	\end{center}
	\medskip
} \fi

\medskip

\begin{abstract}
This paper describes the methodology used by the team \textit{RedSea} in the data competition organized for EVA 2021 conference.
We develop a novel two-part model to jointly describe the wildfire count data and burnt area data provided by the competition organizers with covariates.
Our proposed methodology relies on the integrated nested Laplace approximation combined with the stochastic partial differential equation (INLA-SPDE) approach.  
In the first part, a binary non-stationary spatio-temporal model is used to describe the underlying process that determines whether or not there is wildfire at a specific time and location. 
In the second part, we consider a non-stationary model that is based on log-Gaussian Cox processes for positive wildfire count data, and a non-stationary log-Gaussian model for positive burnt area data.
Dependence between the positive count data and positive burnt area data is captured by a shared spatio-temporal random effect.
Our two-part modeling approach performs well in terms of the prediction score criterion chosen by the data competition organizers.
Moreover, our model results show that surface pressure is the most influential driver for the occurrence of a wildfire, whilst surface net solar radiation and surface pressure are the key drivers for large numbers of wildfires, and temperature and evaporation are the key drivers of large burnt areas.
\end{abstract}
\noindent
{\it Keywords}: INLA-SPDE; Marked point process model; Multivariate processes; Non-stationarity; Spatio-temporal model; Wildfire modeling.

%\spacingset{1.4} % DON'T change the spacing!
%\clearpage

%%%%%%%%%%%%%%%%%%%%%%%%%%%%%%%%%%%%%%%%%%%%%%%%%%%%%%%%%%%%%%%%%%%%%
\section{Introduction}
Wildfires have significant social and economic impact, and might pose significant threat to infrastructure, human safety, and natural resources \citep{Rosenthal2021,Burke2021}.
Furthermore, wildfires are an important source of ${\rm CO}_2$ emissions and contribute substantially to the global greenhouse effect \citep{LiuGoodrick2014}.
Over the past four decades, the wildfire burnt area has roughly quadrupled in the United States (US), which has led to substantial increases in US government expenditures on wildfire suppression in recent years \citep{Burke2021}.
There is thus a pressing need to develop flexible statistical models for wildfire activity and to improve our understanding of wildfire risks so as to support fire management decision-making.

Wildfire risks consist of various components, such as fire occurrence, fire intensity and growth, fire duration, and fire size.
Statistical science has played a key role in modeling and prediction of these components; see \citet{Taylor2013} and \citet{XiTaylor2019} for an overview.
In the data competition of the Extreme Value Analysis (EVA) 2021 conference, the main goal is modeling and prediction of aggregated monthly numbers of wildfire occurrences and their burnt areas in each cell of a regular grid covering the continental US.

Log-Gaussian Cox processes, which are Poisson point processes with intensity specified by a Gaussian random field, have been identified as useful models for wildfire occurrences \citep{Serra2014,Opitz2020}.
As for the burnt area, which represents the size of wildfires, a variety of different models have been considered in the literature, including (truncated) power-law distributions \citep{Cumming2001, Butry2008}, the Weibull distribution \citep{Reed2002}, or the log-normal distribution \citep{Hantson2016}, for positive burnt area data , as well as the generalized Pareto distribution for large wildfires only \citep{Holmes2008}.
The number of wildfires and their burnt areas are clearly linked since if one of them is zero, the other one must also be zero, and a large number of wildfires might correspond to large burnt areas.
It is thus sensible to model these two wildfire characteristics jointly.

A natural approach is to consider a marked point process model with point process identifying the occurrences of wildfires and marks identifying the sizes.
However, the marks might not be separable from the points \citep{Schoenberg2004}, and thus one challenge is how to specify the dependence between them.
Here, we propose a novel two-part model to jointly describe the wildfire count data and burnt area data with environmental covariates, by combining the integrated nested Laplace approximation fast inference with the stochastic partial differential equation (INLA-SPDE) approach.
In the first part, we use a binary spatio-temporal model $Z(\bm{s},t), \bm{s}\in\mathcal{D}, t\in T$, for the underlying process that determines wildfire occurrences, i.e., whether or not there is wildfire at time $t$ and location $\bm{s}$, where $\mathcal{D}\in\mathR^2, T\in\mathR$ are the spatial and temporal domains, respectively.
The first-part modeling is very useful since it accounts for the zero-inflated pattern in the data (more than $60\%$ of the observed count and burnt area data are zeros); see \citet{Liu2019}.
In the second part, we consider a non-stationary log-Gaussian Cox process model $X_{\rm CNT}$ for the shifted positive wildfire count data (minus $1$, specifically), i.e., the point pattern, and a non-stationary Gaussian model $X_{\rm BA}$ for the logarithm of the positive burnt area data, i.e., the marks.
We capture the dependence between the point pattern and marks by a shared spatio-temporal random effect.

The prime goal of this data competition is to estimate the predictive distribution of wildfire occurrences and burnt areas at certain times and sites.
In terms of the prediction score criterion chosen by the data competition organizers, our two-part modeling approach clearly outperforms the benchmark model, which is a generalized linear model with Poisson response for the wildfire count data and a generalized linear model with Gaussian response for the logarithm of positive burnt area data.
Furthermore, we also aim to identify the key drivers for $Z$, $X_{\rm CNT}$, and $X_{\rm BA}$, respectively.

The paper is structured as follows.
Section 2 introduces the data and presents some exploratory analysis.
Section 3 details our modeling approach.
Section 4 presents the results with interpretations.
Section 5 concludes with a discussion.

%%%%%%%%%%%%%%%%%%%%%%%%%%%%%%%%%%%%%%%%%%%%%%%%%%%%%%%%%%%%%%%%%%%%%%%%%%%%
\section{Data}\label{Data}
The dataset contains monthly wildfire information covering March to September from 1993 to 2015 in the continental United States.
More specifically, the study area is partitioned into 3503 cells based on a $0.5^\circ \times 0.5^\circ$ grid of longitude and latitude coordinates.
The number of wildfires and their burnt areas in each grid cell are then aggregated monthly and this yields the two main variables in the dataset, namely counts (CNT) and burnt area (BA).
There are also 35 auxiliary variables, providing the spatial, temporal, meteorological and land cover information.
More details can be found in \citet{Opitz2022}.

\begin{table}[!t]
\centering
\caption{Zero and missing value pattern in the variables CNT and BA of the wildfire dataset}
\begin{tabular}{|l|*{4}{c|}}\hline
\diagbox[width=\dimexpr \textwidth/8+2\tabcolsep\relax, height=1cm]{ BA }{ CNT }
                    & $0$ & $>0$  &  NA & Sum \\ \hline
$0$ & $279762$ & $0$ & $18375$ & $298497$  \\ \hline
$>0$ & $0$ & $173168$ & $12318$ & $185486$ \\ \hline
NA & $18831$ & $12222$ & $48947$ & $80000$  \\ \hline
Sum & $298593$ & $185390$ & $80000$ & $563983$ \\ \hline
\end{tabular}
\label{tab:data_pattern}
\end{table}

Table \ref{tab:data_pattern} shows the zero and missing value pattern in the variables CNT and BA of this dataset.
One can observe that if one of them is zero, the other one must also be zero, and more than $60\%$ of the observed values are zeros.
For statistical modeling of zero-inflated nonnegative continuous data, two different approaches are generally adopted, i.e., a Tobit model or a two-part model \citep{Liu2019}.
The two-part modeling approach is adopted here since it allows us to model the positive wildfire count data and positive burnt areas jointly; see Section \ref{modeling} for more details on the proposed model.

\begin{figure}[!t]
    \centering
    \includegraphics[width=\textwidth]{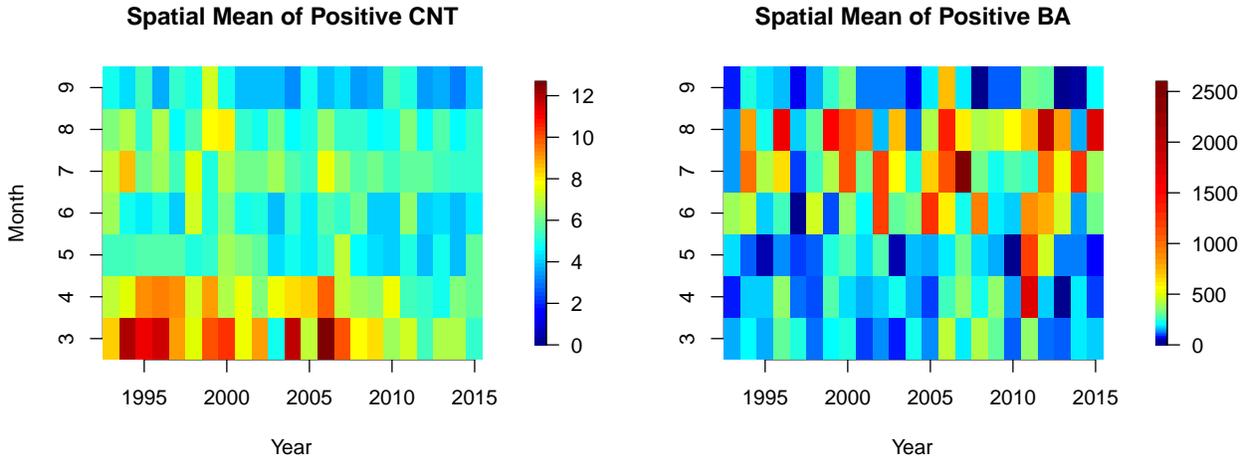}
    \caption{Empirical mean of positive CNTs and BAs, i.e., excluding observations of zeros, in all the grid cells in each month.}
    \label{figure::spatialmean}
\end{figure}

Figure \ref{figure::spatialmean} depicts the monthly empirical mean of positive CNT and BA in all grid cells.
It shows that large wildfire burnt areas often occur for two or three consecutive months, which might be due to the fact that large wildfires often persist for a long period.
Moreover, very large wildfires seldom occur in two nonconsecutive months in the same year due to the reduction of wildland vegetation.
This observation has motivated us to construct spatio-temporal models rather than pure spatial models, aiming to capture the temporal dependence between months.

\begin{figure}[!t]
    \centering
    \includegraphics[width=\textwidth]{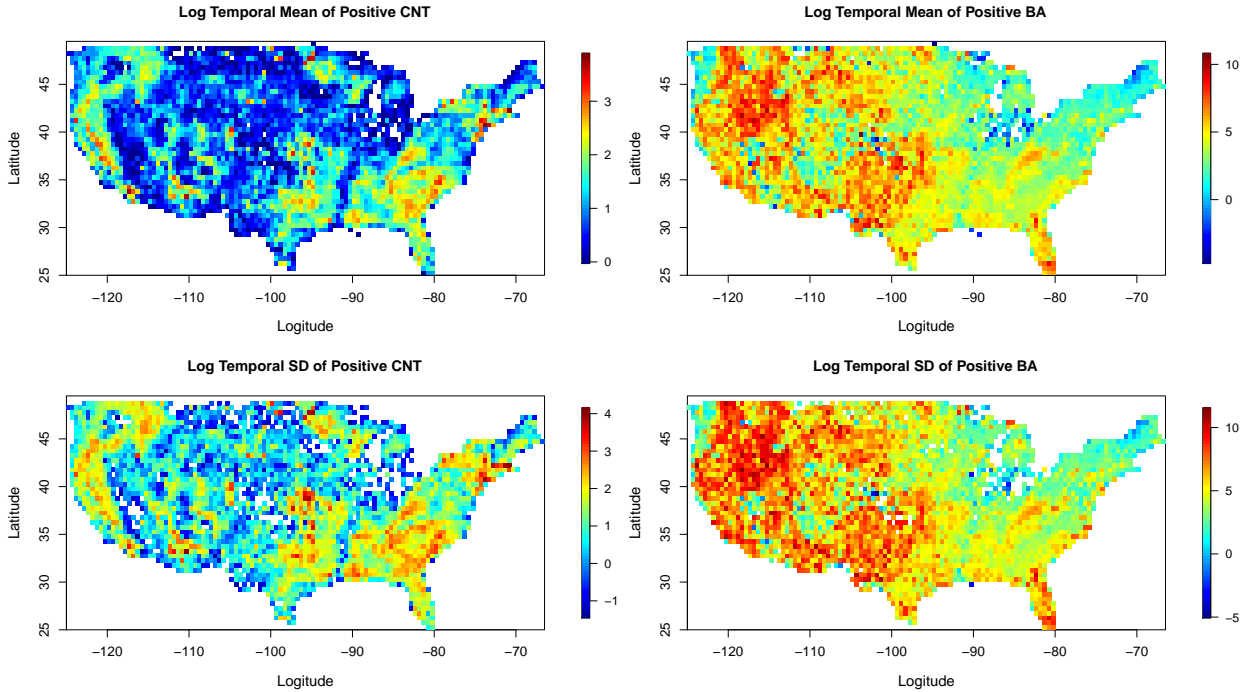}
    \caption{Empirical mean and standard deviation (SD) of positive CNTs and BAs over all months at each grid cell. The grid cells where there are no positive observations are shown in white. The grid cells with only one positive observation, or more than one positive observations but they are equal (resulting in zero SD), are also shown in white in the lower panel.}
    \label{figure::monthmean}
\end{figure}

Figure \ref{figure::monthmean} shows the the logarithm of the empirical mean and standard deviation of the positive CNTs and BAs over time at each grid cell.
One can clearly observe spatial non-stationarity in the mean and standard deviation of the positive observations of both CNT and BA.
Moreover, the pattern in the non-stationarity of CNT and BA appear to be rather different.
Specifically, the western coast and southeastern part of the US appear to have larger numbers of wildfires (large CNTs), but most of the large wildfires (large BAs) appear to occur in the middle- and south-west of the US. 
Motivated by this observation, we propose to use non-stationary spatial models for CNT and BA; see Section \ref{modeling} for details of our method to capture spatial non-stationarity in the mean and standard deviation of CNT and BA.

%%%%%%%%%%%%%%%%%%%%%%%%%%%%%%%%%%%%%%%%%%%%%%%%%%%%%%%%%%%%%%%%%%%%%%%%%%%%
\section{Modeling and Inference}\label{modeling}
\subsection{Two-part Model}
In this section we describe our two-part modeling approach.
Although it is conceptually convenient to think of our constructed models over continuous time and space, we have to discretize the temporal and spatial domain in order to estimate our model in practice.
Our main assumption is that the probability of wildfire occurrence in the binary process, the intensity function of the wildfire point pattern, and the density function of the marks do not vary within each temporal and spatial unit.
Here, the temporal unit is chosen as one month and spatial unit is one grid cell.

The first part of our model is a binary logistic process $Z(\bm{s},t), \bm{s}\in\mathcal{D}, t\in T$, which describes whether or not there is wildfire at site $\bm{s}$ and time $t$.
The sets $\mathcal{D}\in\mathR^2, T\in\mathR$ are the spatial and temporal domains, respectively.
More specifically, our model has the following structure:
\begin{align}
    Z(\bm{s},t) \mid p(\bm{s},t)  &\sim {\rm Bernoulli}(p(\bm{s},t)), \nonumber\\
    {\rm logit}(p(\bm{s},t)) &= \log\big\{\frac{p(\bm{s},t)}{1-p(\bm{s},t)} \big\} = \beta_{0}^Z + \beta_{1}^{Z, {\rm time}} t_{\rm year} + \beta_{2}^{Z, {\rm time}} t_{\rm month} + \nonumber\\
    &\quad\quad\quad\quad\quad \sum_{i=1}^{20} \beta_{i}^{Z, {\rm land}} x_i^{\rm land}(\bm{s}) + \sum_{i=1}^{10} \beta_{i}^{Z, {\rm clim}} x_i^{\rm clim}(\bm{s},t) + \nonumber\\
    &\quad\quad\quad\quad\quad W^Z_1(\bm{s}) + W^Z_2(\bm{s}, a(t)) \label{Zprocess},
\end{align}
where $\beta_0^Z, \beta_i^{Z, {\rm time}}, \beta_i^{Z, {\rm land}}, \beta_i^{Z, {\rm clim}}$ are the regression coefficients to estimate.
There are $32$ covariates in total in the fixed effects, namely $2$ temporal covariates $t_{\rm year} \in\{1993,1994,\dots,2015\}$ and $t_{\rm month}\in\{3,4,\dots,9\}$, $20$ spatial covariates $x_i^{\rm land}, i=1,\dots,20$ including $18$ land cover covariates and $2$ altitude-related covariates, and $10$ meteorological covariates $x_i^{\rm clim}, i=1,\dots,10$.

The process $W^Z_1(\bm{s})$ in (\ref{Zprocess}) is a non-stationary spatial random effect aiming to capture the spatial dependence and non-stationarity in space.
The non-stationarity in $W^Z_1(\bm{s})$ is incorporated in a similar way as \citet{Ingebrigtsen2014} who impose a parametric function of explanatory variables in the parameters that define the Mat\'ern SPDE model.
Here we choose a linear function in terms of the empirical marginal variance as the explanatory variable.
More precisely, we model $W^Z_1(\bm{s})$ through the Mat\'ern SPDE model
\begin{equation*}\label{SPDE}
    \{\kappa^2 - \Delta\}^{\nu+1} \tau(\bm{s}) W^Z_1(\bm{s}) = \dot{\mathcal{M}}(\bm{s}), \quad \bm{s}\in\mathcal{D},
\end{equation*}
where $\nu>0$ is a smoothness parameter and here set to $1$, $\Delta$ is the Laplacian operator, $\dot{\mathcal{M}}$ is Gaussian white noise \citep{Whittle1963}, and where we set
\begin{align*}
    \log\{\tau(\bm{s})\} &= \log(\tau_0) + \theta_1^Z + \theta_2^Z - \hat{\sigma}^Z(\bm{s})\theta_3^Z, \\
    \log\{\kappa\} &= \log(\kappa_0) - \theta_1^Z + \theta_2^Z,
\end{align*}
with $\tau_0,\kappa_0$ constants, $\hat{\sigma}^Z(\bm{s})$ the empirical standard deviation of $Z(\bm{s},t)$ at site $\bm{s}$ over all months, and $\theta_1^Z, \theta_2^Z, \theta_3^Z$ hyperparameters that we need to estimate. 
In this case the solution $W^Z_1(\bm{s})$ is a non-stationary Gaussian random field because $\tau$ varies with location.
Furthermore, the resulting non-stationarity only lies in the marginal variances and we approximately have the marginal variance
\begin{equation*}
    \sigma(\bm{s})^2 \approx \frac{1}{4\pi \kappa^2 \tau(\bm{s})^2}.
\end{equation*}
For more details about the SPDE approach and the construction of non-stationary models, we refer to \citet{Lindgren2011,Ingebrigtsen2014} and \citet{Krainski2019}.

The process $W^Z_2(\bm{s}, a(t))$ in (\ref{Zprocess}) is stationary spatio-temporal random effect defined on the monthly level, i.e., $a(t)\in\{3,\dots,9\}$ denotes the month corresponding to time $t$ and the effects in different years are considered as replicates, aiming to capture the temporal dependence between months and the remaining spatial dependence that is left out by the non-stationary model $W^Z_1(\bm{s})$.
The temporal structure is defined in an autoregressive manner (AR(1), specifically), i.e.,
\begin{equation*}
    W^Z_2(\bm{s}, a(t)) = \rho W^Z_2(\bm{s}, a(t)-1) + \sqrt{1-\rho^2} \epsilon_{a(t)}(\bm{s}), \quad \rho\in(-1,1),
\end{equation*}
for $a(t)\in\{4,5,\dots,9\}$, where $W^Z_2(\bm{s}, 3)=\epsilon_{3}(\bm{s})$, and $\epsilon_{a(t)}(\bm{s})$ are spatial Mat\'ern SPDE innovation fields.

In the second part, we consider modeling the positive observations of CNT and BA jointly.
In order to use a model based on marked point Poisson processes, we subtract the positive CNTs by $1$ to transform the data from range $\{1,2,3,\dots\}$ to nonnegative integers.
Then the resulting point pattern is modeled by a Poisson process $X_{\rm CNT}$ with intensity $\Lambda(\bm{s},t)$, and logarithm of the marks ($\log {\rm BA}$) are modeled by a non-stationary Gaussian process $X_{\rm BA}$.
Specifically, $X_{\rm BA}$ and $\log\Lambda$ have the following additive structures:
\begin{align*}
    X_{\rm BA}(\bm{s},t) &= \beta_{0}^{\rm BA} + \beta_{1}^{{\rm BA}, {\rm time}} t_{\rm year} + \beta_{2}^{{\rm BA}, {\rm time}} t_{\rm month} + \\
    &\quad \sum_{i=1}^{20} \beta_{i}^{{\rm BA}, {\rm land}} x_i^{\rm land}(\bm{s}) + \sum_{i=1}^{10} \beta_{i}^{{\rm BA}, {\rm clim}} x_i^{\rm clim}(\bm{s},t) + \\
    &\quad W^{\rm BA}_1(\bm{s}) + W^{\rm BA}_2(\bm{s}, a(t)) + \epsilon^{\rm BA}(\bm{s},t),\\ 
    \log\Lambda(\bm{s},t) &= \beta_{0}^{\rm CNT} + \beta_{1}^{{\rm CNT}, {\rm time}} t_{\rm year} + \beta_{2}^{{\rm CNT}, {\rm time}} t_{\rm month} + \\
    &\quad \sum_{i=1}^{20} \beta_{i}^{{\rm CNT}, {\rm land}} x_i^{\rm land}(\bm{s}) + \sum_{i=1}^{10} \beta_{i}^{{\rm CNT}, {\rm clim}} x_i^{\rm clim}(\bm{s},t) + \\
    &\quad W^{\rm CNT}_1(\bm{s}) + W^{\rm CNT}_2(\bm{s}, a(t)) + \alpha W^{\rm BA}_2(\bm{s}, a(t)),
\end{align*}
where $\alpha, \beta_0^{\rm BA},\beta_0^{\rm CNT},\beta_i^{{\rm BA}, {\rm type}},\beta_i^{{\rm CNT}, {\rm type}}, {\rm type}\in\{{\rm time},{\rm land},{\rm clim}\}$ are regression parameters to estimate, $\epsilon^{\rm BA}(\bm{s},t)$ can be thought of as a noise or measurement error process which has independent and identical Gaussian distribution with zero mean and unknown precision parameter at any time and space, $W^{\rm CNT}_1(\bm{s})$, $W^{\rm BA}_1(\bm{s})$ are non-stationary spatial random effects with non-stationarity constructed in the same way as $W^{\rm Z}_1(\bm{s})$, and $W^{\rm CNT}_2(\bm{s}, a(t))$, $W^{\rm BA}_2(\bm{s}, a(t))$ are stationary spatio-temporal random effects with AR(1) temporal structures constructed as $W^{\rm Z}_2(\bm{s}, a(t))$.
The dependence between $X_{\rm BA}$ and $\log\Lambda(\bm{s},t)$ is specified by the shared random effect $W^{\rm BA}_2(\bm{s}, a(t))$, and controlled through the parameter $\alpha >0$.
Prior distributions and hyperparameters are discussed in the next section.

\subsection{Bayesian Inference using INLA}
\begin{figure}[!t]
    \centering
    \includegraphics[width=\textwidth]{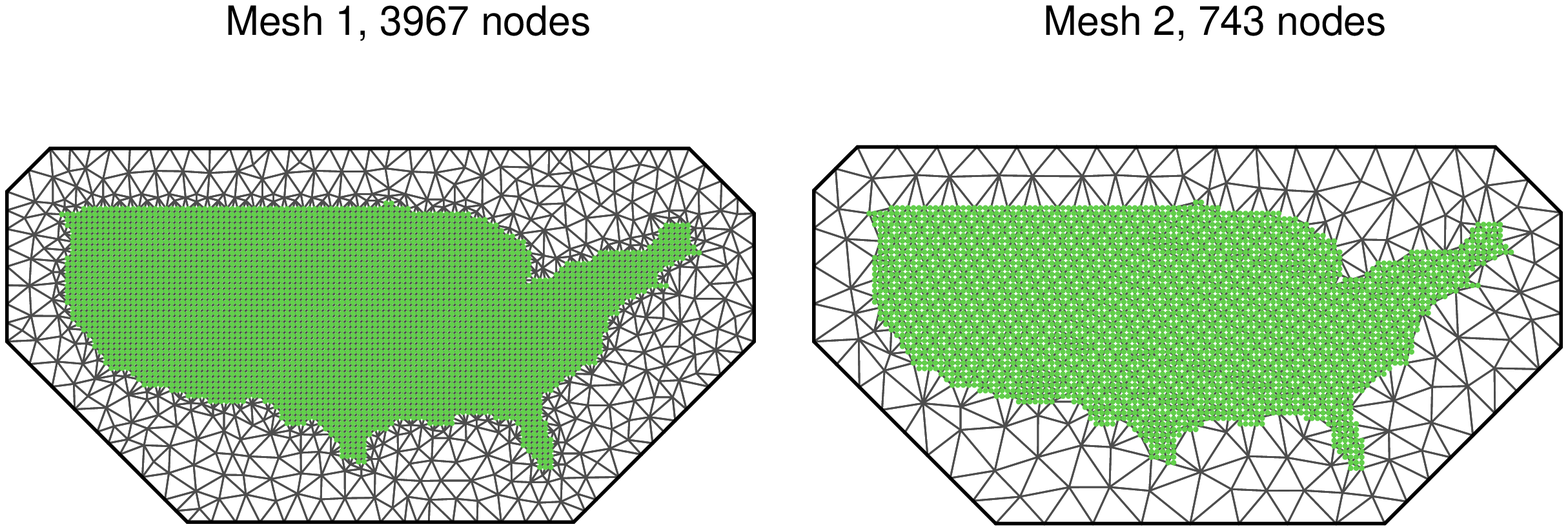}
    \caption{Triangulation over continental US. Green dots indicate the observational sites. Left panel: fine mesh with 3967 nodes for spatial random effects $W_1^j, j\in\{Z, {\rm BA, CNT}\}$; right panel: coarse mesh with 743 nodes for spatio-temporal random effects $W_2^j, j\in\{Z, {\rm BA, CNT}\}$.}
    \label{figure::mesh}
\end{figure}

For each of the random effects $W_i^j, i=1,2, j\in\{Z, {\rm BA, CNT}\}$, we use the SPDE approach to approximate the Gaussian random fields with Mat\'ern covariance by Gaussian Markov random fields, thus enabling computationally efficient inference with INLA \citep{Rue2009, Lindgren2011}.
The SPDE approach is based on a triangulation of the bounded spatial domain $\mathcal{D}$.
Due to computationally considerations, we here choose a fine mesh with 3967 nodes (mesh 1 in Figure \ref{figure::mesh}) for the spatial random effects $W_1^j, j\in\{Z, {\rm BA, CNT}\}$ and a coarse mesh with 743 nodes (mesh 2 in Figure \ref{figure::mesh}) for the spatio-temporal random effects $W_2^j, j\in\{Z, {\rm BA, CNT}\}$.

For the model $Z$, we have $6$ hyperparameters, i.e., $\theta^Z_1, \theta^Z_2, \theta^Z_3$ for the non-stationary spatial random effect $W_1^Z$, and a range parameter, a variance parameter, and the temporal autoregression coefficient $\rho$ for the spatio-temporal random effect $W_2^Z$.
Here we choose a penalized complexity (PC) prior \citep{Simpson2017, Fuglstad2019} for the range parameter, variance parameter, and $\rho$ of random effect $W_2^Z$, and default vague priors in the R-INLA package \citep{Lindgren2015} for other parameters.
Specifically, the PC prior distributions are fixed such that the prior probability of having a covariance range less than $55$ km is $0.1$, of having a variance larger than $0.25$ is $0.1$, and of having temporal autocorrelation parameter $\rho$ below $0$ is $0.05$.
The joint model of $X_{\rm CNT}$ and $X_{\rm BA}$ has $14$ hyperparameters, including one precision parameter for $\epsilon^{\rm BA}$, three parameters for each of $W_i^j, i=1,2, j\in\{\rm BA, CNT\}$, and parameter $\alpha$ for the shared random effect.
We again use PC priors for the temporal autoregression parameter, the range parameter, and variance parameter of the spatio-temporal random effect $W_2^j$ and default vague priors for other parameters.
The PC priors for $W_2^j, j\in\{\rm BA, CNT\}$ are set the same as that for $W_2^Z$.

INLA provides a useful tool for Bayesian modeling and inference for latent Gaussian models.
This is the case for our two-part model as $Z$ is a binomial regression model with logit link, $X_{\rm BA}$ is a Gaussian model, and $X_{\rm CNT}$ is a Poisson regression model with $\log$ link, all of which are conditionally independent of the data level and include latent effects that are jointly Gaussian.
The new package PARDISO \citep{vanNiekerk2021} has enabled parallel computation in INLA and further increased its scalability, which allows us to fit our complex model to this massive wildfire data.
The computation time for fitting our first-part model is around 2 hours on a cluster with 48 cores and 2.9 TB memory, and around 42 hours for fitting our second-part model on the same cluster.
The R code is available at https://github.com/zhongwei-zh/EVA2021-data-competition.

%%%%%%%%%%%%%%%%%%%%%%%%%%%%%%%%%%%%%%%%%%%%%%%%%%%%%%%%%%%%%%%%%%%%%%%%%%%%
\section{Results}

\subsection{Prediction Performance}
We first report the prediction performance of our model as this is the goal of this data competition.
All participants of the competition are required to submit an estimation of the distribution of CNT and BA evaluated at a list of $28$ severity values, at $80000$ different time and locations.
The prediction score of each participant is then calculated based on a modified version of weighted ranked probability score chosen by the organizers, where relatively strong weight is assigned to large values of CNT and BA; see \citet{Opitz2022} for more details.
For our model, the prediction score for CNT is $3498.73$ and the one for BA is $3389.51$, which clearly outperforms the benchmark model whose prediction scores for CNT and BA are $5565.15$ and $4244.36$, respectively.

\subsection{Hyperparameter Estimates}
\begin{table}[!ht]
\centering
\caption{Posterior mean estimates and $95\%$ credible intervals of hyperparamters in the models $Z, X_{\rm CNT}, X_{\rm BA}$.}
\begin{tabular}{lcccc}\toprule
Model & Random effect & Hyperparameter & Estimate & 95\% CI \\
\midrule
$Z$ & $W_1^Z$ & $\theta_1^Z$ & 2.07 & [1.97, 2.17] \\
  & $W_1^Z$ & $\theta_2^Z$ & -1.73 & [-1.84, -1.63] \\
  & $W_1^Z$ & $\theta_3^Z$ & 0.073 & [0.052, 0.093] \\\cmidrule(r){2-5}
  & $W_2^Z$ & Spatial range (km) & 429 & [413, 447] \\
  & $W_2^Z$ & Standard deviation & 1.99 & [1.94, 2.03] \\
  & $W_2^Z$ & Temporal autocorrelation & 0.851 & [0.845, 0.856] \\
\midrule
$X_{\rm CNT}$ & $W_1^{\rm CNT}$ & $\theta_1^{\rm CNT}$ & 2.38 & [2.36, 2.39] \\
  & $W_1^{\rm CNT}$ & $\theta_2^{\rm CNT}$ & -0.83 & [-0.85,- 0.82] \\
  & $W_1^{\rm CNT}$ & $\theta_3^{\rm CNT}$ & -0.33 & [-0.34, -0.31] \\\cmidrule(r){2-5}
  & $W_2^{\rm CNT}$ & Spatial range (km) & 389 & [378, 398] \\
  & $W_2^{\rm CNT}$ & Standard deviation & 0.64 & [0.61, 0.66] \\
  & $W_2^{\rm CNT}$ & Temporal autocorrelation & 0.812 & [0.804, 0.818] \\\cmidrule(r){2-5}
  & Shared random effect & $\alpha$ & 0.703 & [0.696, 0.711] \\
\midrule
$X_{\rm BA}$ & $W_1^{\rm BA}$ & $\theta_1^{\rm BA}$ & 2.05 & [2.03, 2.06]  \\
  & $W_1^{\rm BA}$ & $\theta_2^{\rm BA}$ & -0.52 & [-0.55, -0.50] \\
  & $W_1^{\rm BA}$ & $\theta_3^{\rm BA}$ & 0.089 & [0.082, 0.096] \\\cmidrule(r){2-5}
  & $W_2^{\rm BA}$ & Spatial range (km) & 175 & [171, 179] \\
  & $W_2^{\rm BA}$ & Standard deviation & 1.38 & [1.36, 1.39] \\
  & $W_2^{\rm BA}$ & Temporal autocorrelation & 0.482 & [0.470, 0.494] \\\cmidrule(r){2-5}
  & $\epsilon^{\rm BA}$ & Precision & 0.268 & [0.266, 0.270] \\
\bottomrule
\end{tabular}
\label{tab:hyperparameters}
\end{table}
In this section we report estimates of the hyperparameters in our model.
Table \ref{tab:hyperparameters} presents the posterior mean estimates and $95\%$ credible intervals of all the hyperparameters.
The results show that non-stationarity in the spatial random effects $W_1^{\rm Z}$, $W_1^{\rm CNT}$, and $W_1^{\rm BA}$ are all significant since the $95\%$ credible intervals for $\theta_3^Z$, $\theta_3^{\rm CNT}$ and $\theta_3^{\rm BA}$ do not cover zero.
Temporal dependence between months are very significant since estimates of the temporal autocorrelation parameters for $W_2^{\rm Z}$, $W_2^{\rm CNT}$, and $W_2^{\rm BA}$ are all far from zero and their $95\%$ credible intervals do not contain zero.
Furthermore, spatial dependence for $Z$ and $X_{\rm CNT}$ seems to be stronger than that for $X_{\rm BA}$ as estimates of the range parameters of $W_2^Z$ and $W_2^{\rm CNT}$ are much larger than that of $W_2^{\rm BA}$.
Finally, the dependence between $X_{\rm CNT}$ and $X_{\rm BA}$ is significant and inclusion of the shared random effect is necessary since estimate of the coefficient $\alpha$ is far from zero and its $95\%$ credible interval does not contain zero. 

\subsection{Influence of Covariates on $Z, X_{\rm CNT}, X_{\rm BA}$}
\begin{table}[!ht]
\centering
\caption{Three most positively and negatively influential covariates (based on the values of their posterior means) for the models $Z, X_{\rm CNT}, X_{\rm BA}$. The type of the covariates, the posterior mean estimates of the corresponding coefficients, and their $95\%$ credible intervals are presented.}
\begin{tabular}{lcccc}\toprule
Model & Covariate & Type & Estimate & 95\% CI \\
\midrule
$Z$ & Shrubland & Land cover & -2.67 & [-4.44, -0.91] \\
  & Cropland rainfed herbaceous cover & Land cover & -2.04 & [-3.54, -0.55] \\
  & Grassland & Land cover & -1.76 & [-3.16, -0.37] \\\cmidrule(r){2-5}
  & Surface pressure & Climate & 2.50 & [1.88, 3.12] \\
  & Temperature & Climate & 0.62 & [0.46, 0.77] \\
  & Surface net solar radiation & Climate & 0.56 & [0.50, 0.63] \\
\midrule
$X_{\rm CNT}$ & Shrubland & Land cover & -0.43 & [-0.79, -0.07] \\
  & Dewpoint temperature & Climate & -0.28 & [-0.35, -0.22] \\
  & Cropland rainfed herbaceous cover & Land cover & -0.25 & [-0.55, 0.06]\\\cmidrule(r){2-5}
  & Surface net solar radiation & Climate & 0.40 & [0.36, 0.43] \\
  & Surface pressure & Climate & 0.32 & [0.17, 0.46] \\
  & Temperature & Climate & 0.21 & [0.14, 0.28] \\
\midrule
$X_{\rm BA}$ & Dewpoint temperature & Climate & -0.58 & [-0.71, -0.45] \\
  & Altitude mean & Topography & -0.49 & [-0.82, -0.17] \\
  & Shrubland & Land cover & -0.40 & [-1.54, 0.74] \\\cmidrule(r){2-5}
  & Temperature & Climate & 0.60 & [0.48, 0.73] \\
  & Evaporation & Climate & 0.38 & [0.34, 0.42] \\
  & Surface net solar radiation & Climate & 0.19 & [0.14, 0.25] \\
\bottomrule
\end{tabular}
\label{tab:covariates}
\end{table}

In addition to accurate wildfire prediction, we also aim at identifying the most influential covariates on the three different responses $Z, X_{\rm CNT}$, and $X_{\rm BA}$.
Following the recommendation of \cite{Gelman2008}, we standardize all the covariates in a preliminary step to make them have mean $0$ and standard deviation $1$, so that the effects of different covariates are comparable and interpretable.
Table \ref{tab:covariates} presents the three most positively and negatively influential covariates and their estimated coefficients for the models $Z, X_{\rm CNT}, X_{\rm BA}$.
One interesting observation is that the most influential covariates for the three models $Z, X_{\rm CNT}$, and $X_{\rm BA}$ are in general not the same.
More specifically, surface pressure is the most positively influential covariate to the occurrence of a wildfire, which might be the case because lightning is the principle natural cause of wildfire ignition and surface pressure is often quite high when lightning occurs.
Moreover, while high surface pressure may lead to large numbers of wildfires (possibly because of lightning), high water evaporation, which often occurs when the temperature is high, the air is dry and the wind is strong, leads to large burnt areas.
For the negatively influential covariates, shrubland is a significant covariate for three models, especially for $Z$ and $X_{\rm CNT}$.
This might be due to the fact that large areas of shrubland means less human activity and this results in less human-caused wildfires.
Furthermore, high dew point temperature leads to less wildfires and smaller burnt areas, which might be due to the fact that the higher the dew point temperature, the greater the amount of moisture in the air.
Finally, altitude is a negatively influential covariate for $X_{\rm BA}$, which might be explained by the fact that high altitude corresponds to lower temperature and often less human activity.

%%%%%%%%%%%%%%%%%%%%%%%%%%%%%%%%%%%%%%%%%%%%%%%%%%%%%%%%%%%%%%%%%%%%%%%%%%%%
\section{Discussion}
In this paper we have proposed a novel two-part statistical model for jointly modeling zero-inflated wildfire count data and burnt area data.
Our model clearly outperforms the benchmark model in terms of its prediction performance. Although it performs slightly worse than the model proposed by the best-performing team (named ``BlackBox'') of this data competition, which uses algorithmic models based on deep learning methods, our model yields interpretable results and some understanding of the most important causative drivers that may trigger wildfires.

There are various interesting future research directions.
For instance, one can explore better usage of the covariates information.
Here we only considered a linear additive structure of the temporal, land cover, altitude-related, and meteorological covariates, but a non-linear relationship between some of them and the response variable might exist.
Alternatively, one can investigate how to better build the non-stationary random effects.
Here we chose the empirical marginal variance as the explanatory variable for the variance parameter of the SPDE model, but one can include further variables as one wishes, as long as the added computational cost is acceptable.
One could also consider constructing more complex space-varying regression models as in \citet{OpitzBakka2022}.
All these efforts might be rewarded with a better prediction performance.

Finally, here we did not consider asymptotic models justified by extreme value theory for large values of CNT or BA. 
One reason is that current spatio-temporal extremes models are limited to problems of moderate dimensions and are not suitable for the massive wildfire dataset.
Another practical reason is that although an approach similar to \citet{Opitz2018} or \citet{CatroCamilo2019} may be taken to consider a generalized Pareto distribution for high threshold exceedances of $\log{\rm BA}$, a key feature in the burnt area data is that they are bounded from above, i.e., burnt area cannot exceed the area of their respective grid cells.
This means that the generalized Pareto distribution should be truncated at some point above the threshold, which then brings new modeling challenges and might also involve substantial extra computational cost.
Therefore, investigation of how to integrate extreme value theory in this data application is another interesting future research direction.

%\Appendix
%\appendixpage

%------------------------------------------------------------------------------------
%	                  REFERENCES
%------------------------------------------------------------------------------------

%\small{
\bibliography{reference}
%}

\end{document}